\documentclass[reprint,twocolumn,amsmath,amssymb,aps,pre,floatfix]{revtex4-2}

\usepackage{graphicx}
\usepackage[hypertexnames=false]{hyperref}
\hypersetup{
  colorlinks=true,
  linkcolor=black,
  citecolor=blue,
}
\usepackage{txfonts}
\usepackage{bm}
\usepackage{tabularx}

\newcommand{\kb}{k_{\rm B}}

\begin{document}

\title{Unifying renormalized and bare viscosity in two-dimensional molecular dynamics simulations}
\author{Kazuma Yokota}
\affiliation{Department of Physics, Kyoto University, Kyoto 606-8502, Japan}
\author{Masato Itami}
\affiliation{Department of Physics, Kyoto University, Kyoto 606-8502, Japan}
\author{Shin-ichi Sasa}
\affiliation{Department of Physics, Kyoto University, Kyoto 606-8502, Japan}

\date{\today}
\begin{abstract}
  Fluctuating hydrodynamics provides a framework connecting mesoscopic fluctuations with macroscopic transport behavior.
  To bridge mesoscopic and macroscopic transport from microscopic dynamics, we introduce a wavenumber-dependent viscosity, defined via the equilibrium correlation of time-averaged Fourier components of the fine-grained shear stress field.
  Two-dimensional molecular dynamics simulations reveal its small-wavenumber divergence characteristic of the renormalized viscosity, while its large-wavenumber behavior determines the bare viscosity, thereby establishing a link between mesoscopic and macroscopic transport based on microscopic dynamics.
\end{abstract}
\maketitle

\section{Introduction}
While the motion of microscopic particles is generally highly complex, the macroscopic dynamics of the system is often well captured by the time evolution of a few macroscopic variables.
Hydrodynamics represents one of the most successful frameworks for describing such macroscopic behavior.
Independent of the underlying microscopic interactions, hydrodynamics effectively describes phenomena such as laminar flow, shock waves, and turbulence~\cite{LandauLifshitzFluid1958,Tritton1988}.
These equations are deterministic and neglect the fluctuations inherent at the microscopic level.

Hydrodynamics has recently emerged as a powerful tool for describing a wide range of systems, even extending beyond conventional fluids to include electronic systems~\cite{LucasFong2018,FritzScaffidi2024}, quark-gluon plasmas~\cite{BuszaRajagopalVanderSchee2018}, granular materials~\cite{AransonTsimring2006}, active matter~\cite{Toner2024}, and fracton phases~\cite{GromovLucasNandkishore2020}.
However, in systems of two or fewer dimensions, transport coefficients such as thermal conductivity exhibit system-size-dependent divergence, as observed in experiments~\cite{ChangOkawaGarciaMajumdarZettl2008, XuPereiraWangWuZhangZhaoBaeBuiXieThongHongLohDonaldioLiOzyilmaz2014, LeeWuLowLeeChang2017, YangTaoZhuAkterWangPanZhaoZhangXuChenXuChenMaoLi2021}.
This anomalous behavior is attributed to the contribution of thermal fluctuations.
Therefore, modeling low-dimensional systems requires frameworks that account for these fluctuation contributions beyond deterministic descriptions.

Fluctuating hydrodynamics provides a theoretical framework that incorporates thermal fluctuations into the hydrodynamic equations~\cite{LandauLifshitzFluid1958,Schmitz1988}.
The transport coefficients that appear as parameters in fluctuating hydrodynamics have been referred to as bare transport coefficients since Zwanzig first noted~\cite{Zwanzig1972}.
See also~\cite{ZwanzigNordholmMitchell1972,MoriFujisaka1973} for related papers.
The bare transport coefficients are generally distinct from the macroscopic transport coefficients observed in experiments or numerical simulations, which are interpreted as renormalized transport coefficients resulting from nonlinear fluctuations~\cite{Zwanzig1972,ZwanzigNordholmMitchell1972,MoriFujisaka1973,ForsterNelsonStephen1977,NarayanRamaswamy2002}.
Since the bare transport coefficients govern the behavior of fluctuating hydrodynamics, determining their values is essential for quantitative predictions of transport properties in low-dimensional systems where the standard hydrodynamics is not applicable.
Therefore, the microscopic understanding of the bare transport coefficients is of both practical and fundamental importance.

Given that fluid systems are composed of microscopic particles, determining bare transport coefficients from a microscopic particle description is a significant challenge.
In this context, a formula for the bare transport coefficients in fluctuating hydrodynamics was proposed by~\cite{ZubarevMorozov1983}, based on a projection operator method equivalent to that by Zwanzig~\cite{Zwanzig1961}.
Although formally correct under several technical assumptions, numerically calculating the bare transport coefficients from this formula is intractable.
To overcome this difficulty, a numerically tractable formula combining coarse-graining with the projection operator method was developed~\cite{SaitoHongoDharSasa2021}.
Using this, the bare thermal conductivity was estimated for the one-dimensional Fermi--Pasta--Ulam model, though confirmation of its validity remains a subject for future study.
More recently, nonequilibrium measurements near boundaries have been proposed as a practically useful method~\cite{NakanoMinamiSaito2025}.
However, this estimation method currently lacks a comprehensive theoretical foundation and requires a separate determination of the ultraviolet cutoff.
Despite these advances in understanding bare transport coefficients, a systematic framework for unifying bare and renormalized transport coefficients from a microscopic particle description is still missing.

In this paper, we address this gap by focusing on the bare viscosity $\eta_0$ in a two-dimensional fluid.
In fluctuating hydrodynamics, $\eta_0$ is assumed to be a constant~\cite{LandauLifshitzFluid1958}.
In contrast, the renormalized viscosity, $\eta_{\rm R}(L)$ for a system of size $L$, diverges as $L \to \infty$~\cite{ForsterNelsonStephen1977,NakanoMinamiSaito2025}, highlighting a clear difference between $\eta_{\rm R}(L)$ and $\eta_0$.
Furthermore, viscosity in incompressible fluids is theoretically simpler to analyze than other transport coefficients.
These considerations suggest that connecting $\eta_{\rm R}(L)$ and $\eta_0$ through molecular dynamics simulations offers a crucial first step toward developing a general microscopic theory for bare transport coefficients.

The heart of our approach is to bridge the renormalized viscosity and the bare viscosity through the use of a wavenumber-dependent viscosity $\eta_*(k)$, where $k$ is the magnitude of the wavevector $\bm{k}$.
$\eta_*(k)$ is expected to be independent of the system size $L$ in the thermodynamic limit $L \to \infty$.
The renormalized viscosity $\eta_{\rm R}(L)$ is conventionally computed using the Green--Kubo formula~\cite{Green1954}, which relates it to the correlation of the spatially and temporally averaged fine-grained shear stress field.
Based on this fact, we extract $\eta_*(k)$ from the correlation of time-averaged Fourier components of the fine-grained shear stress field.
By comparing the behavior of $\eta_{\rm R}(L)$ for large $L$ with that of $\eta_*(k)$ for small $k$, we find their equivalence with $L=2\pi \sqrt{2}/k$.
Subsequently, by examining the behavior of $\eta_*(k)$ at large $k$ within the framework of nonlinear fluctuating hydrodynamics, which incorporates $\eta_0$, we determine the value of $\eta_0$ as well as the cutoff length scale $a_{\rm uv}$ of the fluctuating hydrodynamics.

\section{Setup}
We consider a two-dimensional periodic system of $N$ particles confined to the domain $[0,L]\times[0,L]$.
Let $\bm{r}_i$ and $\bm{p}_i$ denote the position and momentum of the $i$-th particle, respectively, and let $m$ be the particle mass.
The Hamiltonian of the system is given by
\begin{align}
  H = \sum_i \frac{|\bm{p}_i|^2}{2m} + \frac{1}{2}\sum_{i \neq j} V(|\bm{r}_i - \bm{r}_j|)
  \label{EQ:Hamiltonian}
\end{align}
with the interaction potential defined as
\begin{align}
 V(r) = \kappa\left(\frac{r_c-r}{\sigma}\right)^2\Theta\left(\frac{r_c-r}{\sigma}\right),
\end{align}
where $\Theta(x)$ is the Heaviside step function.
This potential is chosen to enhance the effects of viscosity renormalization and to facilitate direct comparison with~\cite{NakanoMinamiSaito2025}, which employs the same form of interaction.
Here, $\sigma$ is the particle diameter, and the cutoff distance is fixed at $r_c = 1\sigma$.
The interaction strength is controlled by the parameter $\kappa$, which is set to $\kappa = 10\kb T$, where $\kb$ is the Boltzmann constant and $T$ is the temperature.
All physical quantities are converted to dimensionless forms by setting $\kb T = \sigma = m = 1$.
The density is fixed at $\rho = 0.75$, and the largest system considered consists of $N = 12288$ particles in a periodic box of size $L = 128$.
Molecular dynamics simulations are performed using LAMMPS~\cite{LAMMPS}, where the equations of motion associated with the Hamiltonian given in Eq.~\eqref{EQ:Hamiltonian} are integrated with the velocity-Verlet algorithm~\cite{velocityVerlet} using a time step of 0.001.
To prepare an initial state at temperature $T$, we first equilibrate the system using a Langevin thermostat.
After equilibration, a uniform velocity shift is applied to all particles to ensure that the total momentum is zero.

\begin{figure}[tb]
  \centering
  \includegraphics[keepaspectratio,width=0.85\columnwidth]{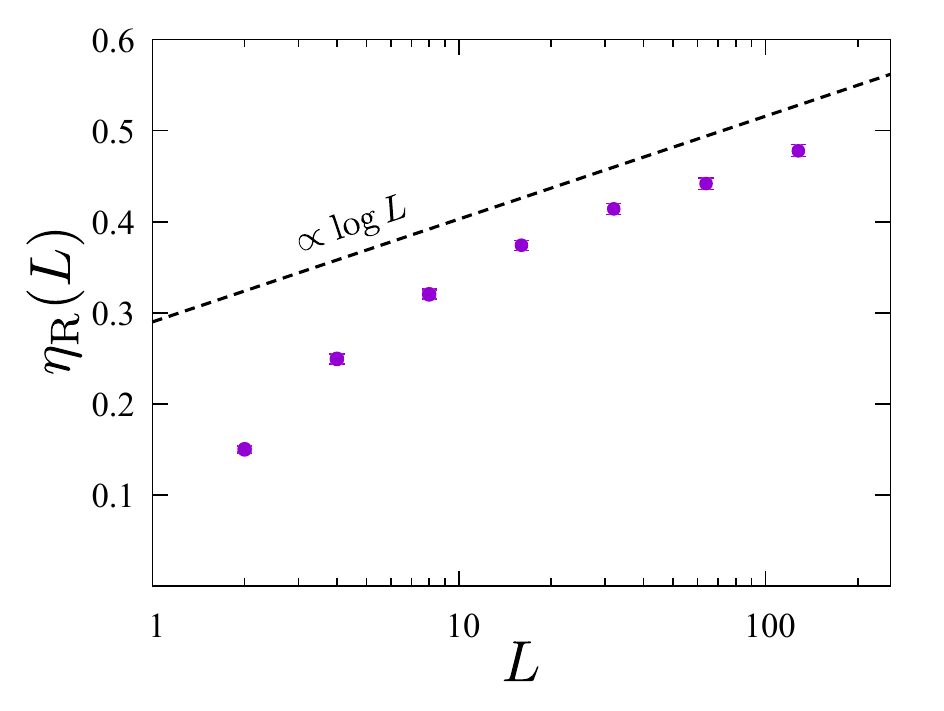}
  \caption{System size dependence of $\eta_{\rm R}(L)$.}
  \label{FIG:System_size_dependence_of_renormalized_shear_viscosity}
\end{figure}

Following the Irving--Kirkwood theory~\cite{IrvingKirkwood1950}, the $a$-component flux of $b$-component momentum at position $\bm{r}$ and time $t$ is defined as
\begin{align}
  \begin{aligned}
    \Pi_{ab}^{\rm IK}(\bm{r},t) &= \sum_i \frac{p_{i,a} p_{i,b}}{m}\delta(\bm{r}-\bm{r}_i) \\
    &- \sum_{i<j} \frac{\partial V(r_{ij})}{\partial r_{ij}}\frac{r_{ij,a} r_{ij,b}}{r_{ij}}\int_0^1 d\lambda\ \delta(\bm{r}-\bm{r}_j-\lambda\bm{r}_{ij}),
    \label{EQ:IK}
  \end{aligned}
\end{align}
where $\bm{r}_{ij}=\bm{r}_i - \bm{r}_j$, and $r_{ij} = |\bm{r}_{ij}|$.
In particular, $\Pi_{xy}^{\rm IK}$ represents the fine-grained shear stress field.
The Green--Kubo formula for the renormalized viscosity $\eta_{\rm R}(L)$ in hydrodynamics is given by~\cite{Green1954}
\begin{align}
  \eta_{\rm R}(L) = \lim_{\tau \to \infty} \eta(L, \tau)
  \label{EQ:Green-Kubo_formula}
\end{align}
with
\begin{align}
  \eta(L, \tau) = \frac{1}{\kb T L^2}\int_0^{\tau} dt \int d\bm{r}\int d\bm{r}^{\prime} \langle\Pi_{xy}^{\rm IK}(\bm{r},t) \Pi_{xy}^{\rm IK}(\bm{r}^{\prime},0) \rangle_{\rm eq},
  \label{EQ:Green-Kubo_integral}
\end{align}
where $\langle \cdot \rangle_{\rm eq}$ denotes the canonical ensemble average at temperature $T$.
The renormalized viscosity exhibits system size dependence.
Figure~\ref{FIG:Green-Kubo} shows the $\tau$ dependence of $\eta(L, \tau)$ for various values of $L$.
See Appendix~\ref{Details_of_numerics} for the details of the numerical calculations.
For $L \geq 32$, a regime in which $\eta(L, \tau)$ increases logarithmically with $\tau$ is observed, reflecting the contribution from the long-time tail~\cite{YamadaKawasaki1967, AlderWainwright1970, WainwrightAlderGass1971, PomeauResibois1975}.
This behavior indicates that $\eta_{\rm R}(L)$ diverges as $L \to \infty$.
Indeed, Fig.~\ref{FIG:System_size_dependence_of_renormalized_shear_viscosity} shows $\eta_{\rm R}(L)$ obtained from $\eta(L, \tau)$ at large $\tau$ in Fig.~\ref{FIG:Green-Kubo}.
It is observed that $\eta_{\rm R}(L)$ diverges logarithmically with $L$.

\begin{figure}[tb]
  \centering
  \includegraphics[keepaspectratio,width=0.85\columnwidth]{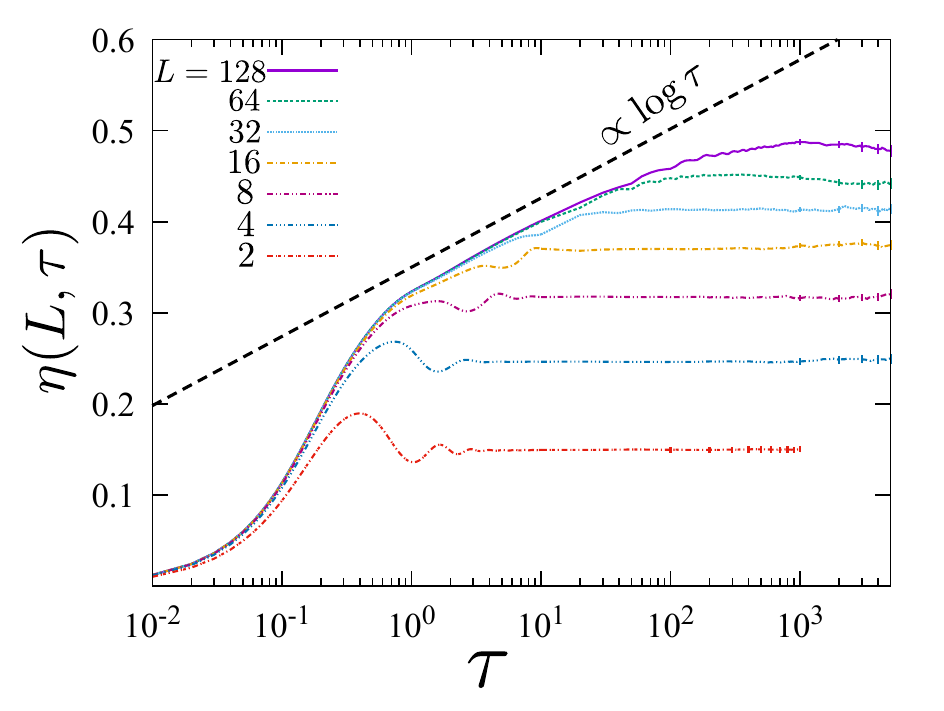}
  \caption{$\tau$ dependence of $\eta(L, \tau)$ for various values of $L$.}
  \label{FIG:Green-Kubo}
\end{figure}

The divergence of $\eta_{\rm R}(L)$ can be understood within the framework of fluctuating hydrodynamics as reviewed in Appendix~\ref{HD_and_FHD}.
For simplicity, we focus on an incompressible and isothermal fluid with density $\rho$
and temperature $T$.
Let $\tilde{v}_a$ denote the $a$-component of the velocity field in Fourier space, restricted to wavevectors $|\bm{k}|\leq 2\pi/a_{\rm uv}$, where $a_{\rm uv}$ is the ultraviolet cutoff that represents the minimal spatial scale at which fluctuating hydrodynamics remains valid.
The time evolution of $\tilde{v}_a$ is governed by the continuity equation
\begin{align}
  \rho\partial_t \tilde{v}_a(\bm{k},t) = - \sum_b i k_b \widetilde{\Pi}_{ab}(\bm{k},t)
  \label{EQ:continuum_equation_for_FHD}
\end{align}
with the momentum flux tensor
\begin{align}
  \begin{aligned}
    \widetilde{\Pi}_{ab}&(\bm{k},t) = \rho \widetilde{v_a v_b}(\bm{k},t) + \tilde{p}(\bm{k},t)\delta_{ab} \\
    &- i \eta_0 \left[k_b \tilde{v}_a(\bm{k},t) + k_a \tilde{v}_b(\bm{k},t)\right] + \sqrt{2\eta_0 \kb T}\tilde{\xi}_{ab}(\bm{k},t),
  \end{aligned}
  \label{EQ:definition_of_momentum_flux_in_FHD}
\end{align}
where the pressure field $\tilde{p}(\bm{k},t)$ is determined from the incompressibility condition $\sum_a k_a\tilde{v}_a(\bm{k},t) = 0$.
The quantity $\widetilde{v_a v_b}(\bm{k},t)$ is the Fourier transform of $v_a(\bm{r},t)v_b(\bm{r},t)$, and $\tilde{\xi}_{ab}(\bm{k},t)$ is a Gaussian white noise satisfying
\begin{align}
  \begin{aligned}
    \langle \tilde{\xi}_{ab}(\bm{k},t) \tilde{\xi}_{cd}(\bm{k}^{\prime},t^{\prime})\rangle &= \Delta_{abcd} L^2\delta_{\bm{k}+\bm{k}^{\prime},\bm{0}}\delta(t-t^{\prime}), \\
    \Delta_{abcd} &= \delta_{ac}\delta_{bd} + \delta_{ad}\delta_{bc} - \delta_{ab}\delta_{cd}.
  \end{aligned}
  \label{EQ:noise_correlation_in_FHD}
\end{align}
The parameter $\eta_0$ in Eq.~\eqref{EQ:definition_of_momentum_flux_in_FHD} is referred to as the bare viscosity~\cite{Zwanzig1972}.
Here, the pole of the velocity correlation function $G_{ij}(\bm{k},\omega)$ in the frequency space provides the renormalized viscosity $\eta_{\rm R}^{\rm FH}(k,\omega)$.
Dynamical renormalization group analysis of fluctuating hydrodynamics~\cite{ForsterNelsonStephen1977} approximately yields
\begin{align}
  \eta_{\rm R}^{\rm FH}(k,\omega=0) \simeq \sqrt{\eta_0^2 - C \rho \kb T\log{\frac{k a_{\rm uv}}{2\pi}}},
  \label{EQ:FNS}
\end{align}
where $C$ is a dimensionless constant given as $1/8\pi$.
Because $k \leq 2\pi/a_{\rm uv}$, we have $\eta_{\rm R}^{\rm FH}(k, \omega = 0) \geq \eta_0$, as follows from Eq.~\eqref{EQ:FNS}.
While a precise relation between $\eta_{\rm R}(L)$ and $\eta_{\rm R}^{\rm FH}(k,\omega=0)$ is not obtained, it is reasonably assumed that the small-$k$ behavior of $\eta_{\rm R}^{\rm FH}(k,\omega=0)$
is equivalent to the large-$L$ behavior of $\eta_{\rm R}(L)$.
Thus, the divergent behavior depending on $\log L$ observed in Fig.~\ref{FIG:System_size_dependence_of_renormalized_shear_viscosity} is well understood by Eq.~\eqref{EQ:FNS}.
See Appendix~\ref{Comparison_with_FNS} for a more detailed comparison.
Putting aside the equivalence of the two quantities, suppose that $\eta_{\rm R}^{\rm FH}(k,\omega=0)$ is measured as a function of $k$ in molecular dynamics simulations.
The important issue is that $\eta_0$ is not determined from $\eta_{\rm R}^{\rm FH}(k,\omega=0)$, because $\eta_0$ is always combined with another unknown parameter $a_{\rm uv}$ as the form $\eta_0^2 - C \rho \kb T\log a_{\rm uv}$.
Thus, determining the bare viscosity from molecular dynamics simulations remains an open problem.

The main purpose of this paper is to propose a formula for determining $\eta_0$ and $a_{\rm uv}$ as intrinsic material parameters that are independent of the system size.
One might conjecture that $\eta_{\rm R}(L_0)$ at some mesoscopic scale $L_0$ provides a bare viscosity.
However, there is no guiding principle to choose $L_0$ from the numerical data in Fig.~\ref{FIG:System_size_dependence_of_renormalized_shear_viscosity}.
A similar problem arises in previous studies identifying a bare transport coefficient based on the concept of scale-dependent transport coefficients~\cite{DonevBellFuenteGarciaPRL2011, DonevBellFuenteGarciaJSM2011}.
In contrast, our formulation naturally connects the renormalized viscosity $\eta_{\rm R}(L)$ and the bare viscosity $\eta_0$ in molecular dynamics simulations, without an ad hoc choice of a mesoscopic scale.

\section{Main results}
A central idea is to extract the $k$-dependent but $L$-independent viscosity ${\eta}_{*}(k)$ defined for the system with large $L$, where $k$ is the magnitude of the wavevector $\bm{k}$.
To obtain ${\eta}_*(k)$, we first define the Fourier transform of the fine-grained shear stress field:
\begin{align}
\widetilde{\Pi}_{xy}^{\rm IK}(\bm{k},t) = \int d\bm{r}\; \Pi_{xy}^{\rm IK}(\bm{r},t)e^{-i \bm{k}\cdot \bm{r}}
\end{align}
for wavevectors $\bm{k} = 2\pi \bm{n}/L$ with $\bm{n} = (n_x, n_y) \in \mathbb{Z}^2$.
Using Eq.~\eqref{EQ:IK}, $\widetilde{\Pi}_{xy}^{\rm IK}(\bm{k},t)$ is expressed as~\cite{CiccottiJacucciMcDonald1979}
\begin{align}
  \begin{aligned}
    \widetilde{\Pi}_{xy}^{\rm IK}(\bm{k},t) = &\sum_i \frac{p_{i,x}p_{i,y}}{m} e^{-i \bm{k}\cdot \bm{r}_i} \\
    &- i \sum_{i>j}\frac{\partial V(r_{ij})}{\partial r_{ij}}\frac{r_{ij,x} r_{ij,y}}{r_{ij}}\frac{e^{-i\bm{k}\cdot \bm{r}_i}-e^{-i\bm{k}\cdot \bm{r}_j}}{\bm{k} \cdot \bm{r}_{ij}}.
  \end{aligned}
\end{align}
Now, a finite-wavevector shear stress correlation is introduced as
\begin{align}
  S(\bm{k},L,\tau) = \frac{1}{2\tau \kb T L^2}\int_0^{\tau}dt \int_0^{\tau}dt^{\prime} \langle \widetilde{\Pi}_{xy}^{\rm IK}(\bm{k},t) \widetilde{\Pi}_{xy}^{\rm IK}(-\bm{k},t^{\prime}) \rangle_{\rm eq},
  \label{EQ:Definition_of_momentum_flux_correlation}
\end{align}
and the $\tau \to \infty$ limit of Eq.~\eqref{EQ:Definition_of_momentum_flux_correlation} is defined as
\begin{align}
  S(\bm{k},L) \equiv \lim_{\tau \to \infty}S(\bm{k},L,\tau),
  \label{EQ:Definition_of_infty_limit_of_momentum_flux_correlation}
\end{align}
where Eqs.~\eqref{EQ:Green-Kubo_formula}, \eqref{EQ:Green-Kubo_integral}, and \eqref{EQ:Definition_of_momentum_flux_correlation} lead to
\begin{align}
  \eta_{\rm R}(L) = S(\bm{k}=\bm{0}, L).
  \label{EQ:consistent}
\end{align}
Note that $S(\bm{k}, L)$ vanishes when $\bm{k}$ is aligned along the $x$- or $y$-axis, as discussed in~\cite{Evans1981}.
This behavior indicates a singularity in the approach to $\bm{k} = \bm{0}$, and does not contradict Eq.~\eqref{EQ:consistent}, since the right-hand side of Eq.~\eqref{EQ:consistent} is defined by first setting $\bm{k} = \bm{0}$ exactly.
Moreover, $S(\bm{k}, L)$ depends on the direction of $\bm{k}$ even for the smallest nonzero wavevectors allowed in a finite system.

\begin{figure}[tb]
  \centering
  \includegraphics[keepaspectratio,width=0.85\columnwidth]{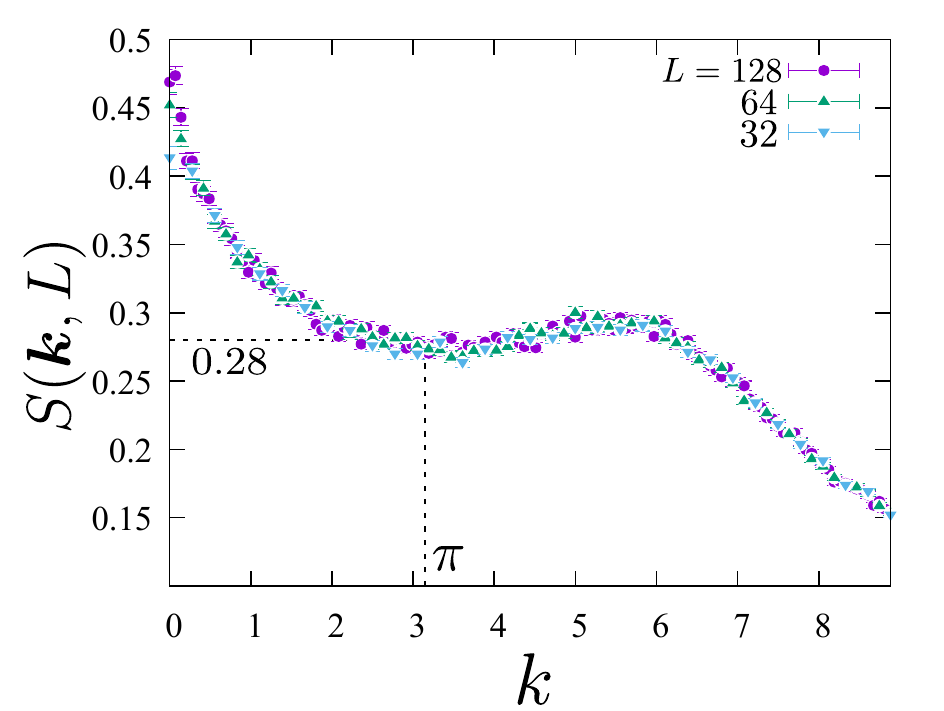}
  \caption{Shear stress correlation $S(\bm{k},L)$ as a function of $k$. We choose $\bm{k}$ such that $k_x = k_y = 2\pi n/L$ with $n \in \mathbb{Z}$, which corresponds to $\bm{k} = (k/\sqrt{2}, k/\sqrt{2})$.}
  \label{FIG:wavenumber_and_systemsize_dependence_of_shear_viscosity}
\end{figure}

A notable feature of $S(\bm{k},L)$ with $\bm{k} \neq \bm{0}$ is that it converges to a limiting function as $L \to \infty$, as confirmed numerically.
Thus, we define $S_\infty(\bm{k})$ as
\begin{align}
  S_\infty(\bm{k}) \equiv \lim_{L \to \infty} S(\bm{k},L) \quad \text{for} \quad \bm{k} \neq \bm{0}.
  \label{EQ:Definition_of_Sinf}
\end{align}
To confirm this behavior, Fig.~\ref{FIG:wavenumber_and_systemsize_dependence_of_shear_viscosity} shows $S(\bm{k},L)$ for $\bm{k} = (k/\sqrt{2}, k/\sqrt{2})$ at three different system sizes.
See Appendix~\ref{Details_of_numerics} for the details of the numerical calculations.
For each nonzero $\bm{k}$, the data for all $L$ collapse onto a single curve, indicating that $S(\bm{k},L)$ is independent of $L$ within numerical accuracy.
This contrasts with the behavior of the conventional viscosity $\eta_{\rm R}(L)$, satisfying $\eta_{\rm R}(L)=S(\bm{k}=\bm{0},L)$, which diverges as $L \to \infty$.

\begin{figure}[tb]
  \centering
  \includegraphics[keepaspectratio,width=0.9\columnwidth]{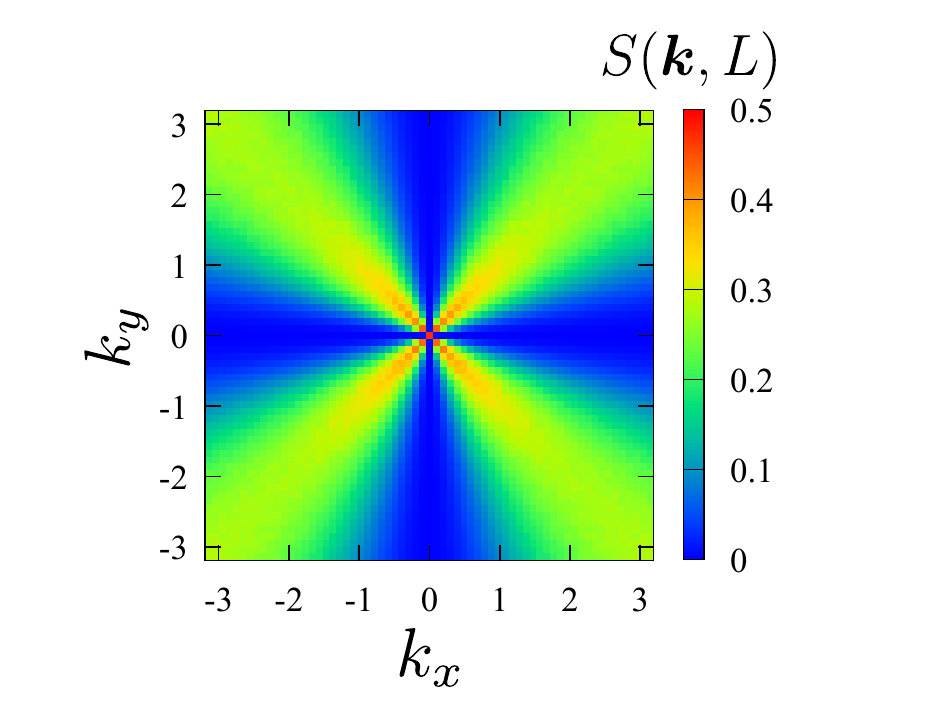}
  \caption{Color display of $S(\bm{k},L=128)$.}
  \label{FIG:Sinf_heatmap}
\end{figure}

Letting $\bm{k} = (k \cos{\theta}, k \sin{\theta})$, we observe that $S_\infty(\bm{k})$ depends on the angle $\theta$.
To extract the $\theta$-independent viscosity $\eta_*(k)$, we focus on the maximum value of $S_\infty(\bm{k})$ over $\theta$:
\begin{align}
  \eta_*(k) \equiv \max_{\theta} S_{\infty}(\bm{k}).
  \label{EQ:definition_of_eta_*}
 \end{align}
It can be shown that this maximum is attained at $\theta = \pi/4$, and that
\begin{align}
  S_\infty(\bm{k}) = \frac{4k_x^2 k_y^2}{k^4}\eta_*(k)
  \label{EQ:Sinf-etas}
\end{align}
for $\bm{k} \neq \bm{0}$.
The key step in the proof of Eq.~\eqref{EQ:Sinf-etas} is to exploit the symmetry properties of the tensor
\begin{align}
  {C}_{abcd}(\bm{k}) \equiv \lim_{\tau \to \infty}\frac{1}{2\tau}
  \int_0^{\tau}dt \int_0^{\tau}dt^{\prime} \langle \widetilde{\Pi}_{ab}^{\rm IK}(\bm{k},t) \widetilde{\Pi}_{cd}^{\rm IK}(-\bm{k},t^{\prime}) \rangle_{\rm eq},
\end{align}
and the identity $\sum_a k_a C_{abcd}=0$, which follows from momentum conservation.
See Appendix~\ref{Derivation_of_Eq17} for a detailed derivation of Eq.~\eqref{EQ:Sinf-etas}.
Note that a similar expression for the shear stress correlation as in Eq.~\eqref{EQ:Sinf-etas} was discussed in viscoelastic fluids~\cite{PicardAjdariLequeuxBocquet2004} and incompressible glasses~\cite{MaierZippeliusFuchs2017}.
To confirm the directional dependence of Eq.~\eqref{EQ:Sinf-etas}, Fig.~\ref{FIG:Sinf_heatmap} shows the $\bm{k}$ dependence of $S(\bm{k},L=128)$.
$S(\bm{k},L)$ exhibits discontinuities near $\bm{k} = \bm{0}$, reflecting its directional anisotropy.
Notably, as $\bm{k}$ approaches $\bm{0}$ along the $\theta = \pi/4$ direction, $S(\bm{k},L)$ converges to $S(\bm{k} = \bm{0}, L)$.
Because $S_{\infty}(\bm{k})$ is equal to $S(\bm{k},L)$ in the limit $L \to \infty$, Fig.~\ref{FIG:Sinf_heatmap} also visualizes the $\bm{k}$ dependence of $S_{\infty}(\bm{k})$ described in Eq.~\eqref{EQ:Sinf-etas}, except at $\bm{k} = \bm{0}$.

\begin{figure}[tb]
  \centering
  \includegraphics[keepaspectratio,width=0.85\columnwidth]{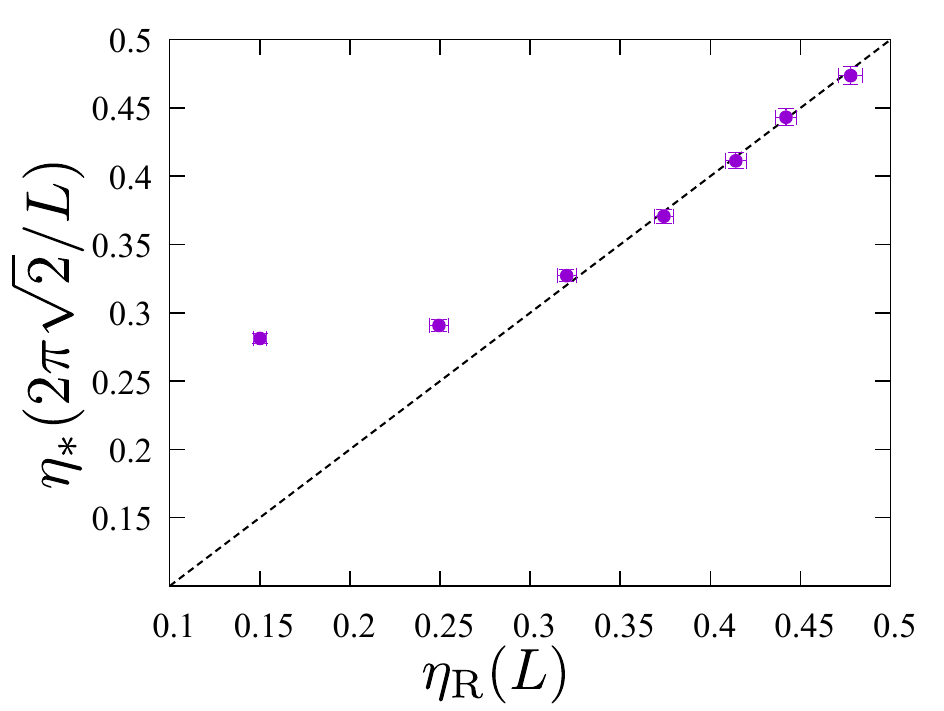}
  \caption{
    Comparison between $\eta_{\rm R}(L)$ and $\eta_*(k=2\pi\sqrt{2}/L)$.
    From left to right, the data points correspond to $L = 2, 4, 8, 16, 32, 64, 128$.
    The dashed line, shown as a guide to the eye, corresponds to $\eta_*(2\pi\sqrt{2}/L) = \eta_{\rm R}(L)$.
    The values of $\eta_*(k)$ are estimated from the simulation data $S(\bm{k},L=128)$.
  }
  \label{FIG:compare_wavenumber_dependent_Green-Kubo_and_general_Green-Kubo}
\end{figure}

The quantity $\eta_*(k)$ defined in Eq.~\eqref{EQ:definition_of_eta_*} is particularly notable for the following two reasons.
First, the large-$L$ behavior of $\eta_{\rm R}(L)$ is closely related to the small-$k$ behavior of $\eta_*(k)$ through the simple relation
\begin{align}
  \eta_{\rm R}(L) \simeq \eta_*(k)
\end{align}
with $k=2\pi\sqrt{2}/L$ for $L \gtrsim 10$, as shown in Fig.~\ref{FIG:compare_wavenumber_dependent_Green-Kubo_and_general_Green-Kubo}.
Second, within the framework of fluctuating hydrodynamics, $\eta_*(k)$ is closely related to the bare viscosity $\eta_0$ through Eq.~\eqref{EQ:Definition_of_momentum_flux_correlation}.
Although an exact calculation is generally difficult, it is reasonable to assume that nonlinear fluctuations can be neglected in the large-wavenumber regime near the ultraviolet cutoff.
Under this assumption, Eq.~\eqref{EQ:definition_of_momentum_flux_in_FHD} without the nonlinear term leads to the explicit expression
\begin{align}
  S_\infty(\bm{k}) = \frac{4k_x^2 k_y^2}{k^4} \eta_0
  \label{EQ:Sinf-eta0}
\end{align}
for $|\bm{k}| \simeq 2\pi/a_{\rm uv}$.
See Appendix~\ref{Derivation_of_Eq20} for details of the calculation.
Note that the same expression is obtained for compressible fluids~\cite{YokotaItamiSasaUnpublished}.
By comparing Eqs.~\eqref{EQ:Sinf-etas} and \eqref{EQ:Sinf-eta0}, we conclude that
\begin{align}
\eta_*(k) = \eta_0,
\end{align}
when $k$ approaches the ultraviolet cutoff $2\pi /a_{\rm uv}$ in fluctuating hydrodynamics.
Several key quantities in this study are summarized in Table~\ref{TAB:Summary_of_viscosity}.

\begin{table*}[tb]
  \centering
  \caption{Summary of viscosity and shear stress correlation in this study.}
  \begin{tabularx}{1.9\columnwidth}{llll}
    \hline
    Notation \hspace{1.5cm}& Equation & Definition \hspace{6cm} & Relation \\
    \hline
    $\eta_{\rm R}(L)$ & \eqref{EQ:Green-Kubo_formula} & $\lim_{\tau \to \infty}\int_0^{\tau} dt \int d\bm{r}\int d\bm{r}^{\prime} \langle\Pi_{xy}^{\rm IK}(\bm{r},t) \Pi_{xy}^{\rm IK}(\bm{r}^{\prime},0) \rangle_{\rm eq}/(\kb T L^2)$ & $\eta_{\rm R}(L) = \eta_*(k)$ with $k=2\pi\sqrt{2}/L$ for $L \gtrsim 10$ \\
    $\eta_0$ & \eqref{EQ:definition_of_momentum_flux_in_FHD} & parameter in fluctuating hydrodynamics & $\eta_0 = \eta_*(k)$ with $k \to 2\pi/ a_{\rm uv}$ \\
    $S(\bm{k},L)$ & \eqref{EQ:Definition_of_infty_limit_of_momentum_flux_correlation} & $\lim_{\tau \to \infty}\int_0^{\tau}dt \int_0^{\tau}dt^{\prime} \langle \widetilde{\Pi}_{xy}^{\rm IK}(\bm{k},t) \widetilde{\Pi}_{xy}^{\rm IK}(-\bm{k},t^{\prime}) \rangle_{\rm eq}/(2\tau \kb T L^2)$ & $S(\bm{k}=\bm{0}, L) = \eta_{\rm R}(L)$\\
    $S_{\infty}(\bm{k})$ & \eqref{EQ:Definition_of_Sinf} & $\lim_{L \to \infty} S(\bm{k},L)$ with $\bm{k} \neq \bm{0}$ & -\\
    $\eta_*(k)$ & \eqref{EQ:definition_of_eta_*} & $\max_{\theta} S_{\infty}(\bm{k})$ & $S_{\infty}(\bm{k}) =  \eta_*(k) 4 k_x^2 k_y^2 / k^4$ \\
    \hline
  \end{tabularx}
\label{TAB:Summary_of_viscosity}
\end{table*}

The graph in Fig.~\ref{FIG:wavenumber_and_systemsize_dependence_of_shear_viscosity} approximately represents $S_\infty(\bm{k})$ with $\bm{k}=(k/\sqrt{2}, k/\sqrt{2})$, which is equivalent to $\eta_*(k)$.
It is observed that ${\eta}_*(k)$ decreases and settles into a plateau for $0 < k \lesssim \pi$, which may correspond to the hydrodynamic regime.
Such a plateau is never observed in $\eta_{\rm R}(L)$.
In the higher-wavenumber region $k \gtrsim \pi$, a slight peak appears, likely reflecting microscopic features of the particle configuration.
These observations suggest that a reasonable choice for the ultraviolet cutoff length is $a_{\rm uv} \simeq 2$, yielding an estimate of $\eta_0 \simeq 0.28$.
Note that, for the same Hamiltonian system with a slightly different density ($\rho = 0.765$), $\eta_0$ and $a_{\rm uv}$ were estimated using nonequilibrium measurements~\cite{NakanoMinamiSaito2025}.
These values, $\eta_0 \simeq 0.3$ and $a_{\rm uv} = \sqrt{2}$ (as defined in our framework), are consistent with our results.

\section{Concluding remarks}
We have formulated the wavenumber-dependent viscosity $\eta_*(k)$ that bridges the renormalized viscosity $\eta_{\rm R}(L)$ and the bare viscosity $\eta_0$.
Because $\eta_*(k)$ is determined by the correlation of the time-averaged Fourier components of fine-grained shear stress fluctuations, the formulation provides the bare viscosity $\eta_0$ and the ultraviolet cutoff $a_{\rm uv}$, which are parameters of fluctuating hydrodynamics, in molecular dynamics simulations. Before ending this paper, we address a few remarks.

The simple method employed in this paper is expected to be applicable to Hamiltonian systems in any dimension for determining other bare transport coefficients.
In particular, for the thermal conductivity, which is frequently measured in experiments~\cite{ChangOkawaGarciaMajumdarZettl2008, XuPereiraWangWuZhangZhaoBaeBuiXieThongHongLohDonaldioLiOzyilmaz2014, LeeWuLowLeeChang2017, YangTaoZhuAkterWangPanZhaoZhangXuChenXuChenMaoLi2021} and numerical simulations~\cite{SaitoHongoDharSasa2021, LepriLiviPoliti2003, Dhar2008}, the bare value is expected to be obtainable by observing the time-averaged Fourier components of the fine-grained heat current defined in~\cite{Sasa2014}.
Beyond conventional fluids, the present framework may also be applicable to systems with order-parameter dynamics~\cite{HohenbergHalperin1977} and active matter~\cite{Toner2024}.

The concept of wavenumber-dependent viscosity has been studied in the theoretical calculation of fluctuating hydrodynamics as $\eta_{\rm R}^{\rm FH}(k, \omega=0)$.
Even in molecular dynamics simulations, this quantity may be numerically evaluated from the decay rate of the momentum correlation function, as studied in~\cite{AlleyAlder1983, Hess2002, HansenDaivisTravisTodd2007, FurukawaTanaka2009}.
We conjecture that $\eta_{\rm R}^{\rm FH}(k, \omega=0)$ is equivalent to $\eta_*(k)$ for $k \leq 2\pi/a_{\rm uv}$.
Numerical investigation of this issue will be reported in another paper~\cite{YokotaItamiSasaUnpublished}.
The microscopic derivation of the relationship among $\eta_{\rm R}(L)$, $\eta_{\rm R}^{\rm FH}(k,\omega=0)$, and $\eta_*(k)$ is an important future subject.

It should be noted that $\eta_{\rm R}^{\rm FH}(k, \omega=0)$ is related to the time correlation function of the projected Fourier mode of the shear stress field~\cite{GrimmBischopinckZumbuschFuchs2024, GrimmBaschnagelSemenovZippeliusFuchs2025}.
Based on the conjecture described in the previous paragraph, this approach can be regarded as obtaining $\eta_*(k)$.
However, unlike our work, the projected dynamics is employed in the time correlation rather than the microscopic dynamics.
While the relation between projected and microscopic stress correlations remains unclear, developing physical arguments for this connection is also an important direction for future investigation.

The finite-wavevector shear stress correlation, Eq.~\eqref{EQ:Definition_of_momentum_flux_correlation}, is the key ingredient in our theory.
Although it is simple and natural, $S(\bm{k},L,\tau)$ has not been extensively studied so far.
The reason is that the finite-wavevector shear stress correlations for wavevectors parallel to certain axes vanish due to momentum conservation~\cite{Evans1981}.
Although this fact was acknowledged in previous studies~\cite{AlleyAlder1983, HansenDaivisTravisTodd2007}, the directional dependence of the finite-wavevector shear stress correlations, as shown in Eq.~\eqref{EQ:Sinf-etas} and Fig.~\ref{FIG:Sinf_heatmap}, does not appear to have been widely recognized.
The quantity $\eta_*(k)$ has been found by noting this directional dependence.

Finally, we discuss potential experimental platforms for probing the bare viscosity.
Two-dimensional fluid behavior has been experimentally observed in systems such as soap films~\cite{Rutgers1998}, dusty plasmas~\cite{NosenkoGoree2004}, and Fermi gases~\cite{VogtFeldFrohlichPertotKoschorreckKohl2012}.
In particular, dusty plasma systems enable control over interaction strength through tunable particle charge, allowing access to regimes where viscosity renormalization becomes significant.
The method proposed in this paper is expected to serve as a tool for determining bare transport coefficients and understanding transport phenomena in such two-dimensional fluids.

\begin{acknowledgments}
The authors are grateful to H. Nakano for sharing his results and for insightful discussions.
The authors also thank M. Fuchs, G. Szamel, A. Yoshida, and R. Suzuki for their helpful discussions, and T. Nakamura for his instruction on LAMMPS.
A part of the computations in this work was performed using the facilities of the Supercomputer Center, the Institute for Solid State Physics, the University of Tokyo.
This study was supported by JSPS KAKENHI Grant Numbers JP22K13975, JP23K22415, JP25K00923, JP25H01975, and JST SPRING Grant Number JPMJSP2110.
\end{acknowledgments}

\section*{Data availability}
The processed datasets used to generate the figures in this manuscript are available on GitHub~\cite{GitHub}.
The corresponding raw simulation data are not publicly available due to their large size but can be obtained from the authors upon reasonable request.

\appendix

\section{Details of the numerical calculations}\label{Details_of_numerics}

In molecular dynamics simulations, the system is equilibrated at $\kb T=1$ using a Langevin thermostat for $10^4$ reduced time units.
After equilibration, the thermostat is removed, and a uniform velocity shift is applied to all particles to ensure zero total momentum.
Subsequently, we perform NVE simulations to measure $\eta(L,\tau)$ and $S(\bm{k},L,\tau)$.
Details of the numerical calculations are described below.

We numerically calculate $\eta(L,\tau)$ by the following procedure for each $(L,\tau)$.
First, $\int d\bm{r}\int d\bm{r}^{\prime} \Pi_{xy}^{\rm IK}(\bm{r},t_0+t) \Pi_{xy}^{\rm IK}(\bm{r}^{\prime},t_0)$ is measured for $0 \leq t \leq t_{\rm max}=5 \times 10^3$ every 0.01 reduced time units ($t_{\rm max}=10^3$ for the $L=2$ system).
The time correlation function with the time interval $t$ is then estimated by the average over $t_0$ satisfying $0 \leq t_0+t \leq 2 \times10^5$.
Finally, these correlations are ensemble-averaged over 512 (1024 for the $L=2$ system) independent initial configurations.
The time integration of the time correlation function over $[0,\tau]$ leads to the results shown in Fig.~\ref{FIG:Green-Kubo}.
The statistical errors are estimated from the unbiased variances of the time integrals of the correlations.

\begin{figure}[tb]
  \centering
  \includegraphics[keepaspectratio,width=0.85\columnwidth]{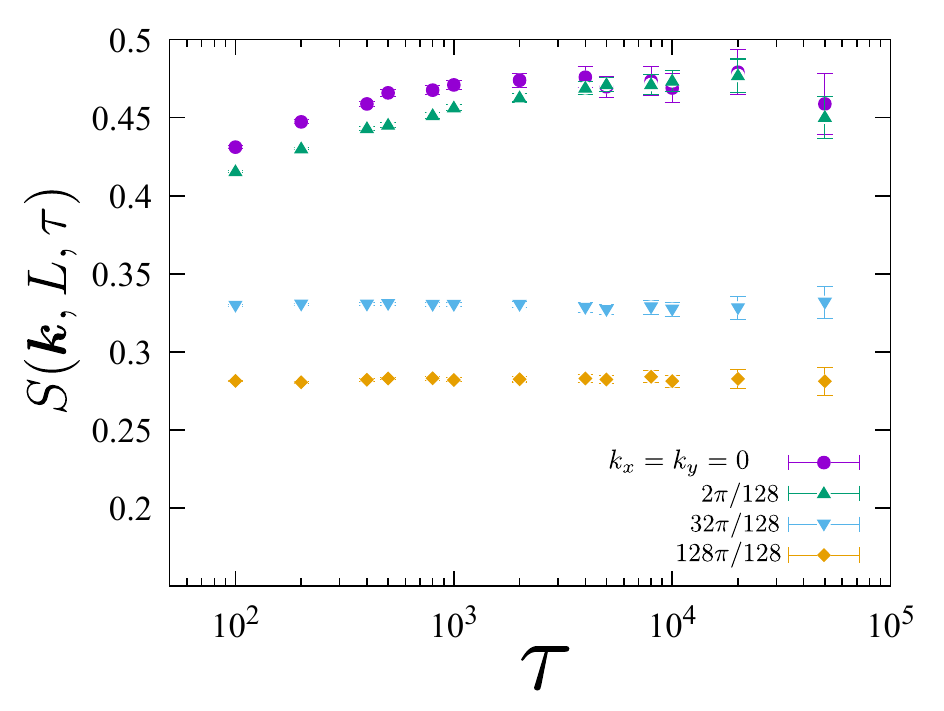}
  \caption{$\tau$ dependence of $S(\bm{k},L=128,\tau)$ for various $\bm{k}$, where $\bm{k}$ satisfies $k_x = k_y$.}
  \label{FIG:correlation_timedep}
\end{figure}

We numerically calculate $S(\bm{k}, L, \tau)$ in Eq.~\eqref{EQ:Definition_of_momentum_flux_correlation} by the following procedure for each $(\bm{k}, L,\tau)$.
First, we calculate $\int_{t_0}^{t_0+\tau}dt \int_{t_0}^{t_0+\tau}dt^{\prime} \widetilde{\Pi}_{xy}^{\rm IK}(\bm{k},t) \widetilde{\Pi}_{xy}^{\rm IK}(-\bm{k},t^{\prime}) / (2\tau \kb T L^2)$ by measuring $\widetilde{\Pi}_{xy}^{\rm IK}(\bm{k}, t)$ every 0.01 reduced time units.
$S(\bm{k}, L, \tau)$ is then estimated by time-averaging the above quantity over $t_0$ satisfying $0 \leq t_0 + \tau \leq 5 \times 10^4$ for each $\tau$.
Finally, we obtain $S(\bm{k}, L, \tau)$ by ensemble-averaging over 1024 independent initial configurations.
The statistical errors are estimated from the unbiased variances of $S(\bm{k}, L, \tau)$.
In Fig.~\ref{FIG:correlation_timedep}, the $\tau$ dependence of $S(\bm{k},L,\tau)$ is shown at some wavevectors.
It is observed that the relaxation time to the long-time limit for smaller $k$ is longer as expected from fluctuating hydrodynamics.
Based on these graphs, we choose $\tau = 10000$ to obtain $S(\bm{k},L)$ from $S(\bm{k},L,\tau)$.

\section{Hydrodynamics and fluctuating hydrodynamics}\label{HD_and_FHD}

In this section, we review the theoretical framework of hydrodynamics and its stochastic extension, fluctuating hydrodynamics.
For simplicity, we consider isolated systems with periodic boundary conditions, where the hydrodynamic equations describe the relaxation from an initially inhomogeneous configuration toward equilibrium, conserving total mass, momentum, and energy.
Fluctuating hydrodynamics introduces stochastic dynamics for the mass, momentum, and energy densities, such that the stationary distribution is consistent with the microcanonical ensemble.

\subsection{Hydrodynamics in three dimensions}

We consider a three-dimensional single-component fluid composed of classical particles evolving under a Hamiltonian.
We assume that the locally conserved quantities---mass density $\rho(\bm{r},t)$, momentum density $\bm{\pi}(\bm{r},t)$, and energy density $h(\bm{r},t)$---constitute the only slow modes of the time evolution.
Each of these quantities satisfies a corresponding continuity equation,
\begin{align}
  \begin{aligned}
    \partial_t \rho + \sum_a \partial_a \pi_a &= 0, \\
    \partial_t \pi_a + \sum_b \partial_b \Pi_{ab} &= 0, \\
    \partial_t h + \sum_a \partial_a J_a &= 0,
    \label{SEQ:continuity_eq_3D}
  \end{aligned}
\end{align}
where $\Pi_{ab}(\bm{r},t)$ is the $ab$-component of the momentum flux tensor,
\begin{align}
  \Pi_{ab}(\bm{r},t) = \rho v_a v_b + \tau_{ab},
  \label{SEQ:momentum_flux_3D}
\end{align}
and $J_a(\bm{r},t)$ is the $a$-component of the energy flux,
\begin{align}
  J_a(\bm{r},t) = (h + p) v_a + \sum_b \tau_{ab} v_b + q_a.
  \label{SEQ:energy_flux_3D}
\end{align}
Here, $v_a(\bm{r},t) = \pi_a(\bm{r},t)/\rho(\bm{r},t)$ is the velocity field, $\tau_{ab}(\bm{r},t)$ is the stress, and $q_a(\bm{r},t)$ is the heat current, defined by
\begin{align}
  \tau_{ab}(\bm{r},t) &= p \delta_{ab} - \eta_{\rm R} \left( \partial_a v_b + \partial_b v_a \right) - \left(\zeta_{\rm R} - \frac{2}{3}\eta_{\rm R} \right) \delta_{ab} \sum_c \partial_c v_c,
  \label{SEQ:stress} \\
  q_a(\bm{r},t) &= - \kappa_{\rm R} \partial_a T.
  \label{SEQ:heat}
\end{align}
The pressure field $p(\bm{r},t)$ and temperature field $T(\bm{r},t)$ are determined from the mass density $\rho(\bm{r},t)$ and the internal energy density $u(\bm{r},t)$, defined as $u = h - \rho |\bm{v}|^2/2$, through thermodynamic relations at each space-time point under the assumption of local equilibrium thermodynamics.
The coefficients $\eta_{\rm R}$, $\zeta_{\rm R}$, and $\kappa_{\rm R}$ in Eqs.~\eqref{SEQ:stress} and \eqref{SEQ:heat} denote the viscosity, bulk viscosity, and thermal conductivity, respectively, and are functions of $\rho$ and $u$.
Because the system consists of interacting particles, these transport coefficients are material-dependent parameters.

The transport coefficients can be expressed as time integrals of equilibrium current-current correlation functions, known as the Green--Kubo formulas.
For instance, under the assumption that the total momentum vanishes and the equilibrium state is spatially uniform, the viscosity is given by
\begin{align}
  \eta_{\rm R} (u, \rho) = \frac{1}{V \kb T} \int_0^{\infty} dt \int d\bm{r} \int d\bm{r}^{\prime} \,
  \langle \Pi_{xy}^{\rm IK}(\bm{r},t) \, \Pi_{xy}^{\rm IK}(\bm{r}^{\prime},0) \rangle_{\rm eq},
\label{GK}
\end{align}
where $V$ is the system volume and $T$ is the temperature.
$\Pi_{ab}^{\rm IK}(\bm{r},t)$ is defined in Eq.~\eqref{EQ:IK} of the main text.
The average on the right-hand side is taken with respect to the equilibrium state specified by $u$ and $\rho$.

From a microscopic perspective, Eqs.~\eqref{SEQ:stress} and \eqref{SEQ:heat} are obtained by a long-wavelength expansion based on the separation between hydrodynamic and microscopic length scales~\cite{Sasa2014}.
Since the transport coefficients are uniquely determined in the thermodynamic limit of the Green--Kubo formulas, the hydrodynamic equations accurately capture the macroscopic time evolution of locally conserved quantities.

\subsection{Fluctuating hydrodynamics in three dimensions}

To describe hydrodynamic fluctuations at equilibrium, thermal noises are introduced so that the detailed balance condition is satisfied.
Explicitly, Eqs.~\eqref{SEQ:stress} and \eqref{SEQ:heat} are replaced by
\begin{align}
  &\begin{aligned}
    \tau_{ab} = &p \delta_{ab} - \eta_0 \left( \partial_a v_b + \partial_b v_a \right) - \left(\zeta_0 - \frac{2}{3} \eta_0 \right) \delta_{ab} \sum_c \partial_c v_c \\
    &+ \sqrt{2 \eta_0 \kb T} \xi_{ab}^{\rm S}+ \sqrt{2 \zeta_0 \kb T} \delta_{ab} \xi^{\rm B},
  \end{aligned}
  \label{SEQ:stress-FH} \\
  &q_a = - \kappa_0 \partial_a T + \sqrt{2 \kappa_0 \kb T^2 }\xi_a^{\rm H},
  \label{SEQ:heat-FH}
\end{align}
where $\xi_{ab}^{\rm S}$, $\xi^{\rm B}$, and $\xi_a^{\rm H}$ are Gaussian white noises satisfying
\begin{align}
  &\begin{aligned}
    &\langle \xi_{ab}^{\rm S}(\bm{r},t) \xi_{cd}^{\rm S}(\bm{r}^{\prime},t^{\prime}) \rangle \\
    &= \left(\delta_{ac} \delta_{bd} + \delta_{ad} \delta_{bc} - \frac{2}{3} \delta_{ab} \delta_{cd}\right) \delta(\bm{r} - \bm{r}^{\prime}) \delta(t - t^{\prime}),
  \end{aligned}
  \label{SEQ:noise-S}\\
  & \langle \xi^{\rm B} (\bm{r},t) \xi^{\rm B}(\bm{r}^{\prime},t^{\prime}) \rangle = \delta(\bm{r} - \bm{r}^{\prime}) \delta(t - t^{\prime}), \\
  & \langle \xi_{a}^{\rm H}(\bm{r},t) \xi_{b}^{\rm H}(\bm{r}^{\prime},t^{\prime}) \rangle = \delta_{ab} \delta(\bm{r} - \bm{r}^{\prime}) \delta(t - t^{\prime}).
\end{align}

The transport coefficients $\eta_0$, $\zeta_0$, and $\kappa_0$ in Eqs.~\eqref{SEQ:stress-FH} and \eqref{SEQ:heat-FH}, which are functions of $(u,\rho)$, are in general distinct from the transport coefficients $\eta_{\rm R}$, $\zeta_{\rm R}$, and $\kappa_{\rm R}$ in Eqs.~\eqref{SEQ:stress} and \eqref{SEQ:heat}.
Macroscopic hydrodynamic behavior can be studied within fluctuating hydrodynamics, for example, through the path-integral formalism.
In this framework, the transport coefficients $\eta_{\rm R}$, $\zeta_{\rm R}$, and $\kappa_{\rm R}$ are obtained as renormalized quantities of $\eta_0$, $\zeta_0$, and $\kappa_0$ due to hydrodynamic fluctuations, analogous to standard field theories.
With this background, it is natural to refer to $\eta_0$, $\zeta_0$, and $\kappa_0$ as the bare transport coefficients, in contrast to the renormalized transport coefficients $\eta_{\rm R}$, $\zeta_{\rm R}$, and $\kappa_{\rm R}$.
See also~\cite{Zwanzig1972,ZwanzigNordholmMitchell1972,MoriFujisaka1973} for further discussion of these concepts.
The subscript ${\rm R}$ of the transport coefficients in the previous section indicates the ``renormalized'' ones.

\subsection{Hydrodynamics and fluctuating hydrodynamics in two dimensions}

Formally, the two-dimensional hydrodynamics is obtained by replacing Eq.~\eqref{SEQ:stress} with
\begin{align}
  \tau_{ab}= p \delta_{ab} - \eta_{\rm R} \left( \partial_a v_b + \partial_b v_a \right) - \left(\zeta_{\rm R} - \eta_{\rm R} \right) \delta_{ab} \sum_c \partial_c v_c.
  \label{SEQ:stress-2d}
\end{align}
However, it is known that the transport coefficients $\eta_{\rm R}$, $\zeta_{\rm R}$, and $\kappa_{\rm R}$ are not determined solely as functions of $(u,\rho)$ but instead depend on system configurations such as the system size.
For example, consider a two-dimensional fluid confined in a square domain $[0,L] \times [0,L]$.
Periodic boundary conditions are imposed in the $x$-direction, and no-slip boundary conditions are imposed at the walls located at $y = 0$ and $y = L$.
We study the simple shear flow generated under the condition that the upper and lower walls move in opposite directions with speed $\dot{\gamma}L/2$, where $\dot{\gamma}$ is the shear rate.
For this simple configuration, the hydrodynamic equations describe the flow with $\eta_{\rm R}$ as a fitting parameter.
In this case, however, unlike in three-dimensional systems, $\eta_{\rm R}$ diverges logarithmically with $L$, as observed in~\cite{NakanoMinamiSaito2025}.
This indicates that $\eta_{\rm R}$ cannot be expressed as a function of $(u,\rho)$ independent of system configurations.
Thus, the hydrodynamic equations cannot be used to predict general hydrodynamic flows in two dimensions.
For this reason, it is often stated that there are no (deterministic) hydrodynamic equations in two dimensions.

From the microscopic viewpoint, this singularity can also be understood via the Green--Kubo formula.
For the simple shear flow illustrated above, $\eta_{\rm R}(L)$, the renormalized viscosity for a system of size $L$, is given by
\begin{align}
  \eta_{\rm R}(L) = \frac{1}{\kb T L^2} \int_0^{\infty} dt \int d\bm{r} \int d\bm{r}^{\prime} \langle \Pi_{xy}^{\rm IK}(\bm{r},t) \Pi_{xy}^{\rm IK}(\bm{r}^{\prime},0) \rangle_{\rm eq}.
\end{align}
Here, due to the long-time tail, the shear stress correlation decays algebraically as
\begin{align}
  \langle \Pi_{xy}^{\rm IK}(\bm{r},t) \Pi_{xy}^{\rm IK}(\bm{r}^{\prime},0) \rangle_{\rm eq} \propto t^{-1}
\end{align}
in the time interval $\tau^{\rm HD} < t < C L^z$, where $\tau^{\rm HD}$ is an $L$-independent onset time of hydrodynamics, $C$ is an $L$-independent constant, and $z$ is a dynamical exponent.
This leads to
\begin{align}
  \eta_{\rm R}(L) = O\left(\int_{\tau^{\rm HD}}^{C L^z} dt \,\frac{1}{t} \right)= O(\log L)
\end{align}
for large $L$.
This result implies that hydrodynamic fluctuations cause the divergence of $\eta_{\rm R}$.
Consequently, it is reasonable to assume that the bare transport coefficients in fluctuating hydrodynamics are given as functions of $(u,\rho)$ independently of system configurations.
With this assumption, one may conjecture that fluctuating hydrodynamic equations exist even in two dimensions, in contrast to the absence of (deterministic) hydrodynamic equations in two dimensions.

Here, fluctuating hydrodynamic equations in two dimensions are obtained by replacing Eqs.~\eqref{SEQ:stress-FH} and \eqref{SEQ:noise-S} with
\begin{align}
  \begin{aligned}
    \tau_{ab}= &p \delta_{ab} - \eta_0  \left( \partial_a v_b + \partial_b v_a \right) - \left(\zeta_0 - \eta_0 \right)  \delta_{ab} \sum_c \partial_c v_c\\
    &+ \sqrt{2 \eta_0 \kb T} \xi_{ab}^{\rm S} + \sqrt{2\zeta_0 \kb T} \delta_{ab} \xi^{\rm B}
  \end{aligned}
  \label{SEQ:stress-FH-2d}
\end{align}
and
\begin{align}
  \begin{aligned}
    &\langle \xi_{ab}^{\rm S}(\bm{r},t) \xi_{cd}^{\rm S}(\bm{r}^{\prime},t^{\prime}) \rangle \\
    &=\left(\delta_{ac} \delta_{bd} + \delta_{ad} \delta_{bc} - \delta_{ab} \delta_{cd}\right) \delta(\bm{r} - \bm{r}^{\prime}) \delta(t - t^{\prime}).
  \end{aligned}
  \label{SEQ:noise-S-2d}
\end{align}
When the energy relaxation is much faster than the momentum relaxation, the equation for the energy density can be replaced by the condition that the temperature is constant.
Furthermore, when the mass density is relatively large, the incompressibility condition may be imposed instead of the explicit time evolution of the mass density.
These are regarded as approximations to the complete fluctuating hydrodynamics and lead to Eqs.~\eqref{EQ:definition_of_momentum_flux_in_FHD} and \eqref{EQ:noise_correlation_in_FHD} in the main text.

Based on the assumption that $\eta_0$ is an $L$-independent parameter of the fluctuating hydrodynamic equations, it follows that the renormalized viscosity $\eta_{\rm R}$ diverges logarithmically as $L \to \infty$~\cite{ForsterNelsonStephen1977}.
Therefore, the existence of fluctuating hydrodynamics in two dimensions is consistent with the absence of deterministic hydrodynamic equations in two dimensions.
The main research question is then to confirm the existence of fluctuating hydrodynamics in two dimensions.
Toward this goal, in the main text, focusing on the viscosity, we discuss how the bare viscosity $\eta_0$ and the spatial cutoff scale $a_{\rm uv}$ in fluctuating hydrodynamics can be determined from the microscopic particle description, as well as how $\eta_0$ is related to the renormalized viscosity $\eta_{\rm R}(L)$.

\section{Comparison of $\eta_{\rm R}(L)$ and $\eta_{\rm R}^{\rm FH}(k)$}\label{Comparison_with_FNS}

\begin{figure}[tb]
  \centering
  \includegraphics[keepaspectratio,width=0.85\columnwidth]{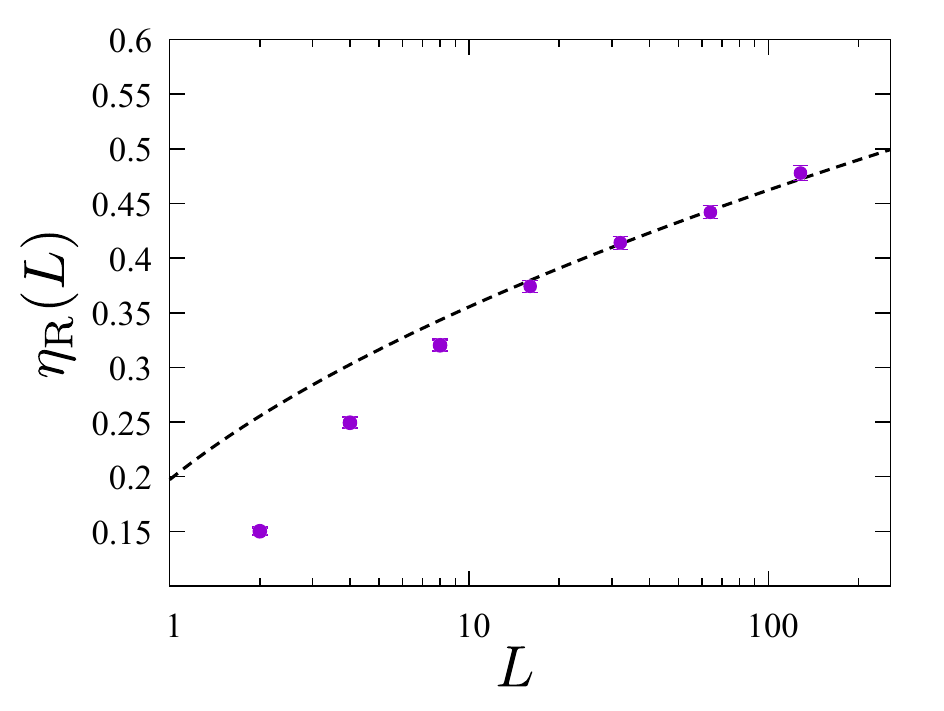}
  \caption{$\eta_{\rm R}(L)$ obtained from molecular dynamics simulations (symbols), together with fitting results using Eq.~\eqref{EQ:FNS} (dashed line).}
  \label{FIG:GreenKubo_compare_FNS}
\end{figure}

In Fig.~\ref{FIG:GreenKubo_compare_FNS}, the system size dependence of the renormalized viscosity $\eta_{\rm R}(L)$ obtained from molecular dynamics simulations is compared with Eq.~\eqref{EQ:FNS}, derived from the dynamical renormalization group analysis of fluctuating hydrodynamics, where we set $k = 2\pi\sqrt{2}/L$.
The dashed line in Fig.~\ref{FIG:GreenKubo_compare_FNS} represents a fit of Eq.~\eqref{EQ:FNS} to the data for $L > 10$, using our estimated value $\eta_0 = 0.28$ and $a_{\rm uv}=2$.
The fitted coefficient is $C = 0.051$, which is slightly larger than the theoretical value $1/8\pi \approx 0.04$ obtained approximately in~\cite{ForsterNelsonStephen1977}.

\section{Derivation of Eq.~\eqref{EQ:Sinf-etas}}\label{Derivation_of_Eq17}

Because the system has no preferred directions, the fourth-rank tensor $C_{abcd}(\bm{k})$ can be generally written as
\begin{align}
  \begin{aligned}
    C_{abcd}(\bm{k}) = &A_1(k) \delta_{ab} \delta_{cd} + A_2(k) \delta_{ac} \delta_{bd} + A_3(k) \delta_{ad} \delta_{bc} \\
    &+ B_1(k) \hat{k}_a \hat{k}_b \delta_{cd} + B_2(k) \hat{k}_a \hat{k}_c \delta_{bd} + B_3(k) \hat{k}_a \hat{k}_d \delta_{bc} \\
    &+ B_4(k) \hat{k}_b \hat{k}_c \delta_{ad} + B_5(k) \hat{k}_b \hat{k}_d \delta_{ac} + B_6(k) \hat{k}_c \hat{k}_d \delta_{ab} \\
    &+ C(k) \hat{k}_a \hat{k}_b \hat{k}_c \hat{k}_d,
    \label{start}
  \end{aligned}
\end{align}
where $\hat{k}_a = k_a / k$. $A_i(k)$, $B_i(k)$, and $C(k)$ are functions of $k=|\bm{k}|$.
Considering symmetry properties $C_{abcd} = C_{bacd} = C_{abdc} = C_{cdab}$, we rewrite Eq.~\eqref{start} as
\begin{align}
  \begin{aligned}
    C_{abcd}(\bm{k}) = &A_1(k) \delta_{ab} \delta_{cd} + A_2(k) (\delta_{ac} \delta_{bd} + \delta_{ad} \delta_{bc}) \\
    &+ B_1(k) (\hat{k}_a \hat{k}_b \delta_{cd} + \hat{k}_c \hat{k}_d \delta_{ab}) \\
    &+ B_2(k) (\hat{k}_a \hat{k}_c \delta_{bd} + \hat{k}_a \hat{k}_d \delta_{bc} + \hat{k}_b \hat{k}_c \delta_{ad} + \hat{k}_b \hat{k}_d \delta_{ac}) \\
    &+ C(k) \hat{k}_a \hat{k}_b \hat{k}_c \hat{k}_d.
  \end{aligned}
\label{e2}
\end{align}
Here, from the continuity equation $\partial_t \tilde{\pi}_a(\bm{k},t) + i\sum_b k_b \widetilde{\Pi}_{ab}^{\rm IK}(\bm{k},t) \allowbreak = 0$, we have
\begin{align}
  \lim_{\tau \to \infty} \frac{1}{\tau}\int_0^{\tau}dt \sum_b k_b \widetilde{\Pi}_{ab}^{\rm IK}(\bm{k},t) = 0,
\end{align}
which yields
\begin{align}
 \sum_b \hat{k}_b {C}_{abcd}(\bm{k}) =0.
\label{e4}
\end{align}
Substituting Eq.~\eqref{e2} into Eq.~\eqref{e4}, we can derive the relations $A_1(k) + B_1(k) = A_2(k) + B_2(k) = B_1(k) + 2 B_2(k) + C(k) = 0$.
Using these relations, we obtain
\begin{align}
  \begin{aligned}
    C_{abcd}(\bm{k}) = &A_1(k) \delta_{ab} \delta_{cd} + A_2(k) (\delta_{ac} \delta_{bd} + \delta_{ad} \delta_{bc}) \\
    &- A_1(k) (\hat{k}_a \hat{k}_b \delta_{cd} + \hat{k}_c \hat{k}_d \delta_{ab}) \\
    &- A_2(k) (\hat{k}_a \hat{k}_c \delta_{bd} + \hat{k}_a \hat{k}_d \delta_{bc} + \hat{k}_b \hat{k}_c \delta_{ad} + \hat{k}_b \hat{k}_d \delta_{ac}) \\
    &+ (A_1(k) + 2 A_2(k)) \hat{k}_a \hat{k}_b \hat{k}_c \hat{k}_d.
  \end{aligned}
\end{align}
Therefore, we have arrived at
\begin{align}
  C_{xyxy}(\bm{k}) = (A_1(k) + 2A_2(k)) \hat{k}_x^2 \hat{k}_y^2.
\end{align}
Noting $S_{\infty}(\bm{k}) = \lim_{L \to \infty} C_{xyxy}(\bm{k}) / (k_{\mathrm{B}} T L^2)$, we next consider the maximum value of $C_{xyxy}(\bm{k})$ over $\theta$ for $k_x=k \cos \theta$ and $k_y=k \sin \theta$.
Because $\hat k_x^2 \hat k_y^2 = \sin^2 (2\theta)/4$, the maximum value is attained at $\theta=\pi/4$.
This yields $\eta_*(k) = \lim_{L \to \infty} [A_1(k) + 2A_2(k)] / (4 k_{\mathrm{B}} T L^2)$, from which Eq.~\eqref{EQ:Sinf-etas} follows.

\section{Derivation of Eq.~\eqref{EQ:Sinf-eta0}}\label{Derivation_of_Eq20}

To simplify the derivation, we define the Fourier transform of a physical quantity $A(\bm{r},t)$ as $\breve{A}(\bm{k},\omega) = \int d\bm{r} \int dt\, A(\bm{r},t) e^{-i\bm{k}\cdot \bm{r}} e^{-i\omega t}$ for the wavevector $\bm{k} = 2\pi \bm{n}/L$ with $\bm{n} = (n_x, n_y) \in \mathbb{Z}^2$.
For Eqs.~\eqref{EQ:continuum_equation_for_FHD}, \eqref{EQ:definition_of_momentum_flux_in_FHD}, and \eqref{EQ:noise_correlation_in_FHD}, we use the Fourier transform and neglect the nonlinear terms.
We then have
\begin{align}
  \rho \omega \breve{v}_a(\bm{k},\omega) = - \sum_b k_b \breve{\Pi}_{ab}(\bm{k},\omega),
\end{align}
\begin{align}
  \begin{aligned}
    \breve{\Pi}_{ab}(\bm{k},\omega) = &- i \eta_0 \left[k_b \breve{v}_a(\bm{k},\omega) + k_a \breve{v}_b(\bm{k},\omega)\right] \\
    &+ \breve{\xi}_{ab}(\bm{k},\omega) - \delta_{ab} \sum_{c,d}\frac{k_c k_d}{k^2} \breve{\xi}_{cd}(\bm{k},\omega),
  \end{aligned}
\end{align}
and
\begin{align}
  \begin{aligned}
    \langle \breve{\xi}_{ab}(\bm{k}, \omega) \breve{\xi}_{cd}(\bm{k}^{\prime}, \omega^{\prime}) \rangle
    &= 2 \eta_0 \kb T 2\pi L^2 \Delta_{abcd} \delta_{\bm{k}+\bm{k}^{\prime},\bm{0}} \delta(\omega+\omega^{\prime}), \\
    \Delta_{abcd} &= \delta_{ac} \delta_{bd} + \delta_{ad} \delta_{bc} - \delta_{ab} \delta_{cd}
  \end{aligned}
\end{align}
with the incompressibility condition $\sum_a k_a \breve{v}_a(\bm{k},\omega) = 0$.
Solving these equations, we obtain
\begin{align}
  \breve{v}_a(\bm{k},\omega) = \frac{\sum_b k_b \breve{\xi}_{ab}(\bm{k},\omega) - k_a \sum_{c,d}\frac{k_c k_d}{k^2} \breve{\xi}_{cd}(\bm{k},\omega)}{\rho \omega - i \eta_0 k^2}.
\end{align}
Using this expression, we calculate
\begin{align}
  \begin{aligned}
    \frac{\langle \breve{v}_a(\bm{k},\omega) \breve{v}_b(\bm{k}^{\prime},\omega^{\prime}) \rangle}{2\kb T 2\pi L^2 \delta_{\bm{k}+\bm{k}^{\prime},\bm{0}} \delta(\omega+\omega^{\prime})}
    &= \eta_0 \frac{k^2 \delta_{ab} - k_a k_b}{\rho^2 \omega^2 + \eta_0^2 k^4}, \\
    \frac{\langle \breve{v}_a(\bm{k},\omega) \breve{\xi}_{bc}(\bm{k}^{\prime},\omega^{\prime}) \rangle}{2\kb T 2\pi L^2 \delta_{\bm{k}+\bm{k}^{\prime},\bm{0}} \delta(\omega+\omega^{\prime})}
    &= \eta_0 \frac{k_c \delta_{ab} + k_b \delta_{ac} - 2 \frac{k_a k_b k_c}{k^2}}{\rho \omega - i \eta_0 k^2}, \\
    \frac{\langle \breve{\xi}_{ab}(\bm{k},\omega) \breve{v}_c(\bm{k}^{\prime},\omega^{\prime}) \rangle}{2\kb T 2\pi L^2 \delta_{\bm{k}+\bm{k}^{\prime},\bm{0}} \delta(\omega+\omega^{\prime})}
    &= \eta_0 \frac{k_b \delta_{ac} + k_a \delta_{bc} - 2 \frac{k_a k_b k_c}{k^2}}{\rho \omega + i \eta_0 k^2}.
  \end{aligned}
\end{align}
We then have the correlation function of the momentum flux:
\begin{align}
  \begin{aligned}
    &\frac{\langle \breve{\Pi}_{ab}(\bm{k},\omega) \breve{\Pi}_{cd}(\bm{k}^{\prime},\omega^{\prime}) \rangle}{2\kb T 2\pi L^2 \delta_{\bm{k}+\bm{k}^{\prime},\bm{0}} \delta(\omega+\omega^{\prime})} \\
    &= - \eta_0^3 k^2 \frac{k_a k_c \delta_{bd} + k_b k_c \delta_{ad} + k_a k_d \delta_{bc} + k_b k_d \delta_{ac}}{\rho^2 \omega^2 + \eta_0^2 k^4} \\
    &\quad - 2 \eta_0 \left( \frac{k_a k_b}{k^2} \delta_{cd} + \frac{k_c k_d}{k^2} \delta_{ab} \right) + \eta_0^3 \frac{4 k_a k_b k_c k_d}{\rho^2 \omega^2 + \eta_0^2 k^4} \\
    &\quad + \eta_0 (\delta_{ac} \delta_{bd} + \delta_{ad} \delta_{bc} + 2 \delta_{ab} \delta_{cd}).
  \end{aligned}
\end{align}
Taking the limit $\omega \to 0$, we obtain
\begin{align}
  \lim_{\omega \to 0} \frac{\langle \breve{\Pi}_{xy}(\bm{k},\omega) \breve{\Pi}_{xy}(\bm{k}^{\prime},\omega^{\prime}) \rangle}{2\kb T 2\pi L^2 \delta_{\bm{k}+\bm{k}^{\prime},\bm{0}} \delta(\omega+\omega^{\prime})} = \frac{4 k_x^2 k_y^2}{k^4}\eta_0,
\end{align}
which becomes equivalent to Eq.~\eqref{EQ:Sinf-eta0} in the limit $L \to \infty$.

\end{document}